\documentclass[8pt]{article}  
\usepackage{times}
\usepackage{graphicx}

\begin{document}

\twocolumn[{\Huge \textbf{The influence of anesthetics,
neurotransmitters and antibiotics on the relaxation processes in lipid
membranes\\*[0.1cm]}}\\

{\large {\center Heiko M. Seeger$^{\dag}$, Marie L. Gudmundsson and Thomas
Heimburg$^{\ast}$\\*[0.1cm] {\normalsize The Niels Bohr Institute,
University of Copenhagen, Blegdamsvej 17, 2100 Copenhagen \O,
Denmark} \\*[0.3cm]}

{\normalsize In the proximity of melting transitions of artificial and
biological membranes fluctuations in enthalpy, area, volume and
concentration are enhanced.  This results in domain formation, changes
of the elastic constants, changes in permeability and slowing down of
relaxation processes.  In this study we used pressure perturbation
calorimetry to investigate the relaxation time scale after a jump into
the melting transition regime of artificial lipid membranes.  This
time corresponds to the characteristic rate of domain growth.  The
studies were performed on single-component large unilamellar and
multilamellar vesicle systems with and without the addition of small
molecules such as general anesthetics, neurotransmitters and
antibiotics.  These drugs interact with membranes and affect melting
points and profiles.  In all systems we found that heat capacity and
relaxation times are related to each other in a simple manner.  The
maximum relaxation time depends on the cooperativity of the heat
capacity profile and decreases with a broadening of the transition.
For this reason the influence of a drug on the time scale of domain
formation processes can be understood on the basis of their influence
on the heat capacity profile.  This allows estimations of the time
scale of domain formation processes in biological membranes.\\*[0.1cm]
}} {\footnotesize \textbf{keywords: }lipid membrane, phase separation,
domain formation, rafts, relaxation, pressure perturbation
calorimetry\\*[0.1cm]} {\footnotesize \textbf{abbreviations:} DSC,
differential scanning calorimetry; PPC, pressure perturbation
calorimetry; DMPC, 1,2-dimyristoyl-sn-glycero-3-phosphocholine; DPPC,
1,2-dipalmitoyl-sn-glycero-3-phosphocholine; halothane,
3-bromo-2-nchloro-1,1,1-trifluoroethane; serotonin,
5-hydroxytryptamine; MLV, multilamellar vesicle; LUV, large
unilamellar vesicle\\*[0.1cm] $^{\ast}$corresponding author,
theimbu@nbi.dk, {http://membranes.nbi.dk}\\
$^{\dag}$present address:National Center of CNR-INFM S3, NanoBioLab,
Via G. Campi 213/A, 41100 Modena, Italy\\*[0.3cm]}]

\section*{Introduction}

Artificial \cite{Steim:dscmembr,Chapman:liqcrypropphoslip,
Hinz:dscbilayer} as well biological membranes \cite{Steim:dscmembr,
melchior:transbiomembr, Heimburg2005c} display melting transitions.  In
their vicinity fluctuations in lipid area, volume, enthalpy and
concentration are enhanced.  As a result domains form, the elastic
constants are increased \cite{heim:mech98, Ebel:enthvolchang} and the
permeability is higher \cite{papa:permeab, Nagle:permeab,
Cruzeiro:permeability,Antonov2005}.  A whole variety of molecules like
proteins, peptides or other small molecules influence melting
transitions.  Nowadays biological membranes are seen as being rather
heterogenous \cite{Jacobson:revfluidmosaic} instead of being
homogenous fluids as suggested in the fluid-mosaic model
\cite{sing:fluidmos}.  Heterogeneities of biological membranes due to
lipid domain formation have been postulated for many years
\cite{Sackmann:triggprocmembrstruct}.  In the biology community,
however, domain formation has only gained a stronger interest since
the discussion about `rafts' \cite{Brown:rafts, Simons:liprafts}.
Rafts are thought to be involved into trafficking processes
\cite{Helms:raftstrafficking}.  Domain formation in general is
believed to trigger biochemical reaction cascades and to influence
enzyme activity \cite{Melo:domainconnectionreactions, Vaz:phasetopol,
Thompson:domainstrucreaction, Hinderliter:controlsigntrans,
Salinas:changesenzymeactivity}.

Domains can form as a consequence of melting transitions.  Already in
the 1970s melting transitions were measured in artificial
\cite{Chapman:liqcrypropphoslip, Hinz:dscbilayer} and in biological
membranes \cite{melchior:transbiomembr, Jackson:EcoliDSC} using
differential scanning calorimetry.  The increased fluctuations in the
melting regime of membranes are accompanied by changes in the
time-scales of the domain formation process.  Studies on the kinetics
of phase transitions in artificial membranes mainly use temperature
and pressure jump techniques \cite{traeuble:schaltprozmemb,
Tsong:kin_bilay, Tsong:relaxphen,
Gruenewald:kininvest,Elamrani:presjumrelax,Holzwarth:krittrueb,
Blume:dmchol_pjrelax, Grabitz2002}, but also volume perturbation
\cite{Johnson:volumepert, vanOsdol1989, vanOsdol1991a, vanOsdol1991b,
vanOsdol1992}, pH or ion concentration changes
\cite{strehlow:kinelctrigg}, ultrasonic measurements
\cite{mitaku:ultrasonicrelax, Halstenberg:CritFluct} and
ac-calorimetry \cite{Yao:relax} were applied.  The number of
relaxation processes in different studies varies from one to five.
Relaxation times in single-lipid membranes were reported to lie in the
range from $ns$ to $min$ \cite{holzwarth:structdyn_PCmembr}, depending
on the experiment and the detection method.  Most studies agreed on a
slowing down of relaxation processes in the transition regime
\cite{traeuble:schaltprozmemb, Tsong:relaxphen,
mitaku:ultrasonicrelax, Gruenewald:kininvest, Elamrani:presjumrelax,
holzwarth:structdyn_PCmembr, vanOsdol1989, Grabitz2002,
Halstenberg:CritFluct, vanOsdol1991a}.  It was also pointed out that
maximum relaxation times were observed at the transition midpoint.  In
\cite{Grabitz2002} relaxation processes were connected to
macroscopical fluctuations and it was shown that the relaxation times
of the cooperative processes and the heat capacity are connected in a
simple manner.

In earlier studies, an influence of cholesterol, dibucaine and
peptides was found \cite{holzwarth:structdyn_PCmembr,
Blume:dmchol_pjrelax, vanOsdol1992, Grabitz2002}, but
not systematically investigated.  Molecules like general anesthetics,
neurotransmitters and antibiotics display a functional role in
biological cells.  The antibiotic gramicidin A is a hydrophobic,
channel-forming peptide \cite{Gennis:biomembranes}.  Dimers of this
peptide induce channels of an outer diameter of about $5\: \AA$ that
is mainly permeable for monovalent cations.  Neurotransmitters are
molecules which mainly occur in the nerve system, but can also be
found in other parts of the body.  They influence nerve pulse
propagation and are either inhibitory or excitatory.  An example of an
excitory neurotransmitter is serotonin (5-hydroxytryptamine).
Neurotransmitters have been discussed to also act as anesthetics
\cite{Cantor:neurotransanesth}.  Anesthesia is the state when pain,
consciousness or other sensations are blocked.  General anesthetics
such as 1-octanol or halothane
(2-bromo-2-chloro-1,1,1-trifluoroethane) lead to a reversible complete
loss of consciousness and sensation.  The action of general
anesthetics is still not fully known.  Theories favoring an direct
influence on protein function are at the moment favored
\cite{Franks:genanaesth, Krasowski:genanesth, Bovill:meyeroverton},
but lipid membrane mediated mechanisms are also discussed.  Ueda and
Yoshida \cite{Ueda:lipmembanesth} claim that anesthetics action
results from effects on both proteins and lipids and they consider
the lipid/water interface.  Cantor \cite{Cantor:genanesthetics,
Cantor:meyeroverton} relates anesthetic function to the influence of
anesthetics on the lateral pressure profile of lipid membranes.
Heimburg and Jackson \cite{Heimburg2007b, Heimburg2007c} attribute
anesthetic action to the influence of anesthetics on the lipid
membrane state.  

Peptides, neurotransmitters and anesthetics all display significant
influence on lipid melting transition.  This immediately raises
questions about these molecules on the macroscopic membrane
properties.  In this paper, we focussed on the influence of
anesthetics, neurotransmitters and antibiotics on the time scales of
domain formation processes.  This is an extension of a previous study
by Grabitz and collaborators \cite{Grabitz2002}.  Relaxation
processes are related to the cooperative fluctuations.  More
specifically, it was shown that heat capacity and relaxation times in
pure lipid membranes are proportional functions.  Here, the study is
extended towards lipid systems with the incorporation of drugs.  It
will be shown that the addition of these molecules does not change the
linear relation between heat capacity and relaxation times, but
systematically alter the details of the relaxation process.  We
discuss the biological importance of our findings in relation to the
influence of domains on biochemical reaction cascades, enzyme
activity, sorting and trafficking in biological membranes.

\section*{Materials and Methods}

\textbf{Sample preparation: }Lipids were purchased from Avanti Polar Lipids (Birmingham/AL, USA)
and used without further purification.  All lipid samples were
dissolved in a $10\:$ $mM$ Hepes buffer with $1\: mM$ EDTA at pH $7$.
Samples of multilamellar vesicles (MLVs) were prepared by adding
buffer to the lipid powder and stirring the solution above the main
phase transition temperature using a magnetic stirrer for at least one
hour.  During this time the lipid solution was vortexed at least three
times.  Suspensions of large unilamellar vesicles (LUVs) were prepared
using an Avestin extruder system (Avestin Europe GmbH, Mannheim,
Germany).  Suspensions of MLVs were extruded at $320.9\: K$ using a
filter with a pore size of 100nm.  DMPC/ neuro\-trans\-mitter systems were
made by adding serotonin (Hydrochloride; Sigma-Aldrich Inc., St.
Louis/MO, USA) to the buffer before dissolving the lipids in the
buffer/neuro\-trans\-mitter solution.  In the case of DMPC/anesthetics
solutions 1-octanol (Sigma-Aldrich Inc., St.  Louis/\-MO, USA) or
halothane (2-Bromo-2-chloro-1,1,1-trifluoroethane, Sig\-ma-Al\-drich Inc.,
St.  Louis/\-MO, USA) was added to the already prepared suspension of
MLVs.  The solution was stirred for another 30 $min$.  DMPC/\-peptide
membrane systems were prepared by dissolving the DMPC lipids and
gramicidin A as powder in organic solvent (dichlo\-romethane: methanol
2:1).  After mixing both solutions the lipid/gra\-micidin A solution was
dried through heating and a light nitrogen or air stream.  The sample
was kept in a high vacuum desiccator over night.  The rest of the
preparation equaled the one of preparing suspensions of MLVs.

\textbf{Calorimetry:} Differential scanning calorimetry (DSC)
measurements were performed with a VP-DSC from Microcal
(Nor\-thhampton/MA, USA) using high feedback mode at a scan rate of
$5 K/hr$ if not indicated otherwise.  A concentration of $10\: mM$
was used to measure the excess heat capacity profile of a sample
directly in the calorimeter cell and determining the transition
enthalpy.  In other cases sample solutions were filled into self-built
pressure cells (see below).  In these cases concentrations of either
$40\: mM$ or $50\: mM$ were prepared.  The capillary volume was not
exactly known, but with the knowledge of the transition enthalpy we
could determine absolute excess heat capacity values.  Heat capacity
profiles of DMPC/gramicidin A mixtures were determined with scan rates
of $1\: K/hr$.  Due to hysteresis effects the scan rate has an
influence on the temperature of the maximum heat capacity.  Curves
were corrected correspondingly.  Sample solutions and buffer solutions
were degassed for at least $15\: min$ before the calorimetric
experiments to avoid measurement artifacts arising from possibly
evolving gas bubbles.
\begin{figure*}[htb]
    \begin{center}
	\includegraphics[width=15cm]{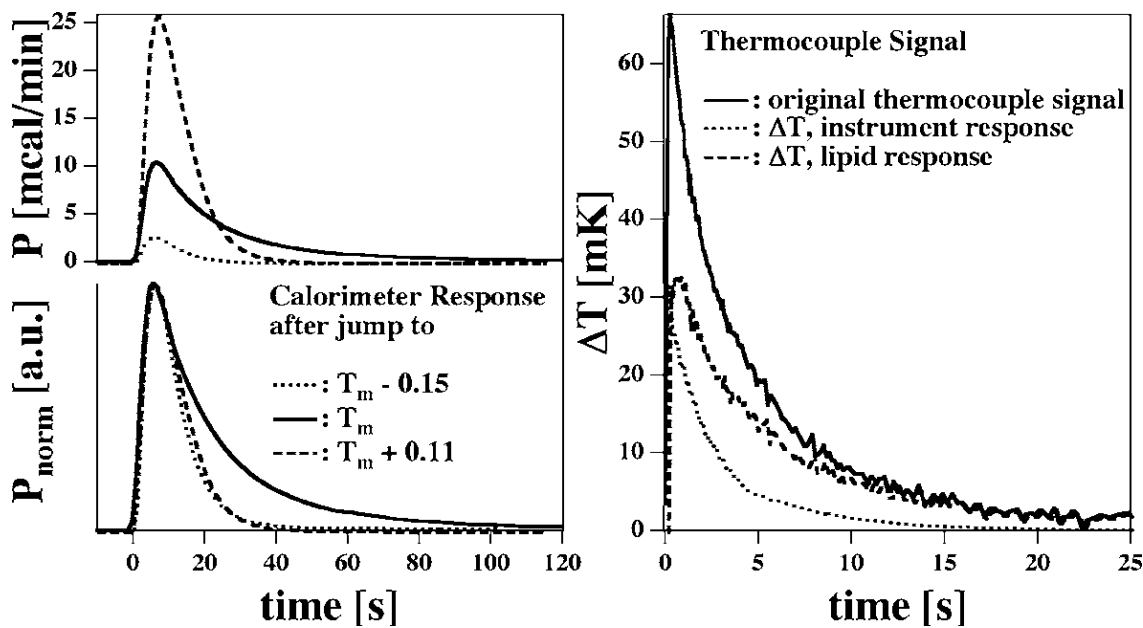}
	\parbox[c]{16cm}{ \caption{\textit{Left: Representative
	calorimeter responses after a negative pressure jump.  The
	curves were obtained after jumps to different temperatures
	within the lipid melting transition of DMPC vesicles.  The
	further one jumps through the transition (for negative
	pressure jumps the higher the temperature) the bigger is the
	integrated heat absorption of the lipid membrane (\emph{left,
	upper panel}).  After normalization of the curves it can be
	seen that they decay on different time scales (\emph{left,
	lower panel}).  Right: Representative temperature change after
	a positive pressure jump recorded by a thin thermocouple.  The
	total response is the sum of the response of the
	water filled capillary (instrument response) and the response 
	from the lipid sample (lipid response).  }
	\label{fig:fig1}}}
    \end{center}
\end{figure*}

\textbf{Pressure perturbation calorimetry:} The same calorimeter was
used for pressure perturbation calorimetry (PPC) experiments in the
isothermal mode of the calorimeter.  Measurements on solutions of MLVs
were performed as previously described \cite{Grabitz2002}.  The
sample solution was filled into a self-built pressure capillary which
can hold pressures up to $500\: bar$.  This capillary was inserted
into the sample cell of the calorimeter.  The melting transition
regime shifts upon addition of hydrostatic pressure by about 1K/40bar
(see \cite{Mountcastle:anestheandpressure} and
\cite{Ebel:enthvolchang} with citations therein).  This is used to
perturb the system.  Pressure is controlled through addition or
release of high pressure nitrogen by opening and closing manual or
solenoid pressure valves (Nova Swiss, Effretikon, Switzerland).  The
change of pressure occurs on a time scale of around $100\: ms$.
Changing the pressure on the lipid dispersions by $\pm 40\: bar$ and
an appropriate choice of temperature allowed to jump to different
points of the phase coexistence regime where domains started to form.
Jumps into the transition regime were either from below the
transition ($-40\: bar$; negative pressure jump) or from above the
transition ($40\: bar$; positive pressure jump)e.
In the first case the signal is endothermic, while in the other case
it is exothermic.  After a pressure change the calorimeter needs to
compensate the heat absorption or release to keep the sample
temperature constant.  This response contains information about the
relaxation times of the domain formation process.  Representative
calorimetric responses after negative pressure jumps ($-40\: bar$) are
shown in the left panel of fig.~\ref{fig:fig1} (top).  The curves are
results after jumps to different points of the transition defined by
different sample temperatures.  The area below the curves, i.e. the
total heat absorption of the lipid membranes during the equilibration
process increases the further one jumps through the transition (here:
increasing temperature).  The normalized signals show that the
relaxation process occurs on different time scales
(fig.~\ref{fig:fig1}, bottom).  A sample cell containing buffer
solution also results in a characteristic response.  This is the
response function of the instrument.  The observed calorimetric signal
from the equilibration process is the convolution of the real signal
(called the ``sample response'') with the instrument response function.
The response after a pressure jump always contained one small
component that contain the perturbation of the pressure cell itself
and small contributions from lipid membrane processes faster than the
resolution of the instrument.  The dominant part of the signal
consists of a single exponentially decaying heat absorption or release
from the lipid membranes after a perturbation.  Faster relaxation
processes only contribute to a minor degree to the overall signal (see
discussion concerning Fig.  6).  The above method allows a time
resolution of about $3-4\: s$.

The relaxation times of lipid dispersions are related to the heat
capacity.  Broader melting profiles display smaller heat capacity
values in the transition.  Under these conditions faster relaxation
processes are found (Grabitz2002) for which the resolution of $3-4\:
s$ may not be good enough.  For this reason, a refined experimental
setup with increased time resolution was used especially for
measurements on LUV dispersion that display melting peaks that are about
10 times broader than the MLV transitions.  We only performed positive
pressure jumps.  The pressure release or addition was controlled by
two computer controlled solenoid valves (Nova Swiss, Effretikon,
Switzerland) and the time scale of the pressure relaxation was faster
than $90\: ms$.  The pressure cell consisted of two capillaries
instead of only one.  A K-thermo\-cou\-ple with a grounded hot junction of
a diameter of $0.25\: mm$ (Thermocoax, Stapelfeld, Germany) was put
into the sample capillary of the pressure cell.  Relaxation times were
determined from a direct measurement of changes in temperature of the
sample solution after the pressure jump.  The thermocouple signal was
amplified by a Nanovolt Preamp Model 1801 (Keithley Instruments Inc.,
Cleveland/OH, USA) and recorded by a Keithley Multimeter 2001
(Keithley Instruments Inc., Cleveland/OH, USA).

As in the case analyzing the calorimeter response the total
temperature change is seen as a convolution of the ``instrument
response'' and the ``sample response''.  This is displayed in the right
panel of fig.~\ref{fig:fig1}.  The total temperature change (solid
curve) after a positive pressure jump can be divided into a change due
to the ``instrument response'' (dotted curve) and the ``lipid
response'' (dashed curve).  The latter one is again modeled by a
convolution of a single exponentially decaying heat release of the
lipid membrane and the instrument response.  For details we refer to
\cite{Grabitz2002, Seeger:PhDthesis2006}.  The contribution from the
sample cell (only filled with water) was subtracted from the total
temperature change and only the remaining signal was analyzed.

Pressure perturbation calorimetry experiments were always conducted on
a few successive days.  Melting profiles were controlled after each
measuring day.  During these days heat capacity curves sometimes
broadened slightly or shifted towards higher or lower temperatures.
The shift could be easily corrected later.  Below it will become
obvious that a broadening of the curve means that relaxation times are
influenced.  The broadening, however, was small so that the
determination of relaxation times was in the range of experimental
error.


\section*{Theory}

In an earlier paper from our laboratory  a theory 
of the relaxation times of lipids in the melting regime was derived 
on the grounds of non-equilibrium thermodynamics \cite{Grabitz2002}. 
This theory made use of the experimental fact that enthalpy, volume 
and area are proportional functions in this temperature regime. 
Therefore, perturbation of the lipid samples by temperature, bulk or 
lateral pressure results in the same relaxation process. It was found 
that one expects a proportional relation between relaxation time and 
excess heat capacity.

The theory is based on the assumption that the distribution of
enthalpy states at a given temperature can be described by a Gaussian
distribution.  This is correct when the system is a continuous transition
that is neither of first order nature nor at a critical point:
\begin{equation}
    P(H-\langle H\rangle)=\frac{1}{\sigma
    \sqrt{2\pi}}e^{-\frac{(H-\langle H\rangle)^2}{2\sigma^2}},
\end{equation}
where $H$ is the enthalpy, $\langle H\rangle$ is the mean enthalpy and
$\sigma^2=\langle H^2\rangle-\langle H\rangle^2$ is the variance.  The
Gibb's free energy depends logarithmical on this distribution
\cite{lee:fssMCs91}: $G=RT\,\ln P$.  Using this one finds that the
entropy $S$ can be approximated as a harmonic potential:
\begin{equation}
    S(H-\langle H\rangle)\approx-\frac{R(H-\langle H
    \rangle)^2}{2\sigma^2}, 
    \label{eq:entropyharmonicpot}
\end{equation} 
where $R$ is the gas constant. 

Throughout a melting transition changes in area, volume and enthalpy
are proportional and, therefore, one only finds one independent
fluctuation given by $\alpha=H-\langle H\rangle$.  The thermodynamic
force driving the lipid sample back to equilibrium (after a
perturbation) is given by $X=d S/d (H-\langle H\rangle)$, and the flux
of heat by $J=d (H-\langle H\rangle)/d t$.  This is in fact the
observable in our calorimetric experiments.  According to Onsager
\cite{Onsager1931a,Onsager1931b} the flux is proportional to the
thermodynamics force.  Therefore
\begin{equation}
    \frac{d(H-\langle H\rangle)}{dt}=-L\frac{R(H-\langle 
    H\rangle)}{\sigma^2})
    \label{eq:phenEq}
\end{equation} 
using the phenomenological constant L. This leads to a simple
differential equation that is solved by an exponential decay
\begin{equation}
    (H-\langle H\rangle)=(H-\langle H\rangle)_{0}\cdot \exp(-\frac{t}{\tau})
    \label{eq:relax}
\end{equation} 
introducing the relaxation time $\tau$.  Using the identity
$c_{p}=(H-\langle H\rangle)^2/RT^2\equiv\sigma^2/RT^2$ (fluctuation
theorem) one arrives at
\begin{equation}
    \tau=\frac{T^2}{L}c_p,
    \label{eq:relaxtipropcp}
\end{equation}   
where $T$ is the temperature and $L$ is a phenomenological constant.
The constant L has to be obtained from experiments. For the details of the
derivation we refer to \cite{Grabitz2002}.  Note, however, that in
the cited paper a factor of $RT$ was omitted in the derivation so that
an unit error occurred.  This, however, does not change the message of
the previous paper that relaxation times and heat capacity are
proportional functions.


\section*{Results}

The non-equilibrium thermodynamics theory briefly outlined above
predicts a proportionality between heat capacity and relaxation times
for the cooperative processes with large excess heat capacity.
Important for this prediction is the cooperative nature of the process
rather than the chemical composition of the membrane.  In
\cite{Grabitz2002} it is shown for different one component lipid
systems (MLV and LUV) and a DMPC/cholesterol mixture (MLV) that one
finds a linear relationship between heat capacity and relaxation times
in pressure perturbation calorimetry.  Maximum relaxation times are up
to about a minute and they are decreased by the addition of
cholesterol in a manner closely related to the effect of cholesterol
on the heat capacity profile.  In the present study we extended the
range of the membrane systems.  In particular, we investigated the
influence of small drugs on domain formation processes and the related
relaxation times.  For this it was necessary to improve the time
resolution of our calorimetric relaxation measurements.  The latter
part is important for large unilamellar membrane systems that display
smaller heat capacities and broader transitions than multilamellar
preparations.
\begin{figure*}[htb]
    \begin{center}
	\includegraphics[width=15cm]{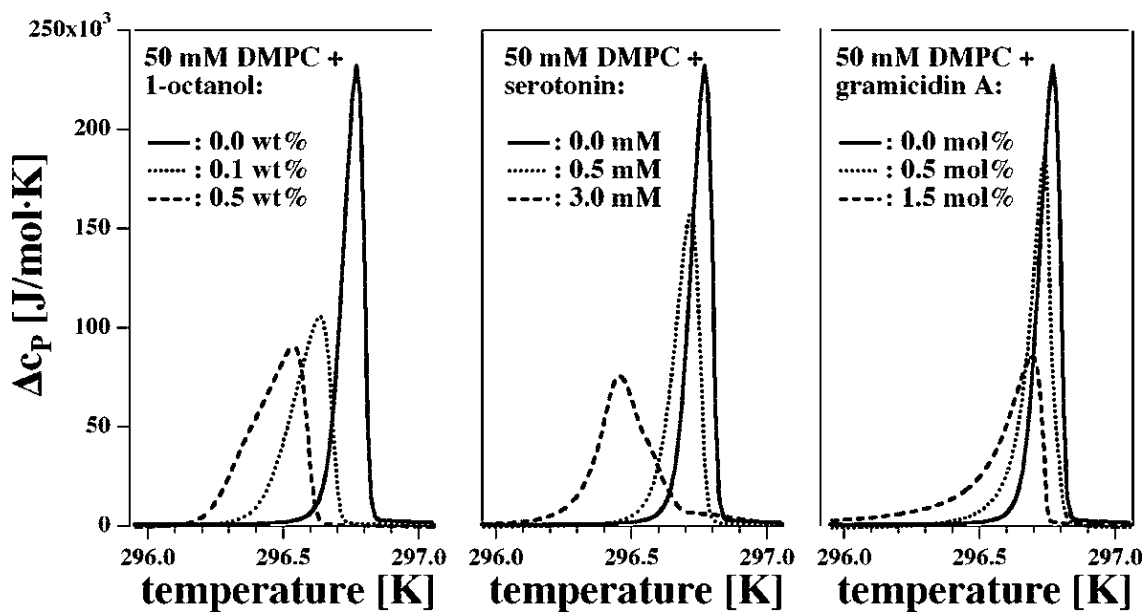}
	\parbox[c]{16cm}{ \caption{\textit{Heat capacity profiles of
	various multilamellar lipid dispersions in the presence of
	drugs.  Left: DMPC plus various concentrations of 1-octanol.
	Center: DMPC plus different concentrations of serotonin Right:
	DMPC plus different concentrations of gramicidin A. In all
	three panels the drugs broaden the lipid melting profile and
	shift it to lower temperatures.}
	\label{fig:fig2}}}
    \end{center}
\end{figure*}

First, we determined the influence of molecules such as general
anesthetics, neurotransmitters and antibiotics on the melting behavior
of lipid membranes by differential scanning calorimetry.  In
fig.~\ref{fig:fig2} the heat capacity profiles of $50\: mM$
multilamellar DMPC vesicles with different concentrations of 1-octanol
(\emph{left panel}), serotonin (\emph{center panel}) and gramicidin A
(\emph{right panel}) are displayed.  The transition enthalpy of these
systems was determined to be $21.5\: kJ/mol$.  The general finding is
that the addition of the respective molecules broadened the transition
profile and shifted it to lower temperatures.  The total melting
enthalpy remains unaltered.  In the following we wanted to understand
in which way relaxation processes were influenced in the respective
systems.

\begin{figure}[b!]
    \begin{center}
	\includegraphics[width=8cm]{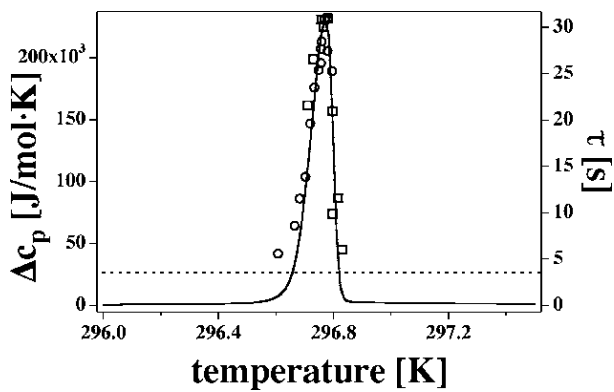}
	\parbox[c]{8cm}{ \caption{\textit{Heat capacity curve
	(\emph{solid curve}) of a $50\: mM$ multilamellar DMPC sample
	and relaxation times (\emph{open circles and squares})
	obtained from pressure perturbation experiments on the same
	sample as measured by the calorimeter.  The different markers
	refer to different experimental situations where pressure is
	released (\emph{open circles}) or added (\emph{open squares}).
	The change in relaxation times correlates nicely with the heat
	capacity profile.  The dashed line indicates the lower
	resolution limit of the experiment (3-4 seconds).  }
	\label{fig:fig3}}}
    \end{center}
\end{figure}

In fig.~\ref{fig:fig3} heat capacity profiles and relaxation times of
a $50\: mM$ DMPC (MLV) solution are displayed.  The heat capacity
curve shown is the same as in fig.~\ref{fig:fig2}.  Relaxation times
were determined from measurements using the same lipid dispersion as
for the DSC scans.  The melting transition profile of lipid membranes
with increased pressure was shifted to higher temperatures, whereas
the shape of the profile was not influenced.  Therefore, the
relaxation time profile obtained from positive pressure jumps were
temperature corrected with $\Delta T=-0.9\: K$ (open
squares).  Results from negative pressure jumps are given by open
circles.  The time resolution of the setup is indicated by a dashed
line. Relaxation times faster than this experimental response time 
cannot be determined accurately.

Comparison of heat capacity and relaxation times shows the linear
relation of these two functions.  In the proximity of the melting
transition relaxation processes slows down.  The maximum relaxation
time measured is about $(31.0\pm1.5)\: s$.  With a maximum heat
capacity value of $232\: kJ/mol\: K$ and the temperature at the
transition midpoint of $296.77\: K$ we found a phenomenological
constant (see eq.~\ref{eq:relaxtipropcp}) of $(6.6\pm 0.5)\cdot 10^8\:
J\: K/mol\: s$.  The previous experiment was performed while recording
the heat compensation of the calorimeter (with a time resolution of
about 4 seconds).  Performing experiments on a new suspension of $40\:
mM$ DMPC (MLV) and evaluating the temperature change with the
thermocouple inserted into the pressure cell (with a time resolution
of about 0.3 seconds) after a positive pressure jump of $+15\: bar$
yielded a phenomenological constant of $(6.8\pm0.3 )\cdot 10^8\:
J\: K/mol\: s$.  Both phenomenological constants obtained
agree within error with each other.  This demonstrates that our two
different setups record the same time scales.

\begin{figure}[htb!]
    \begin{center}
	\includegraphics[width=8cm]{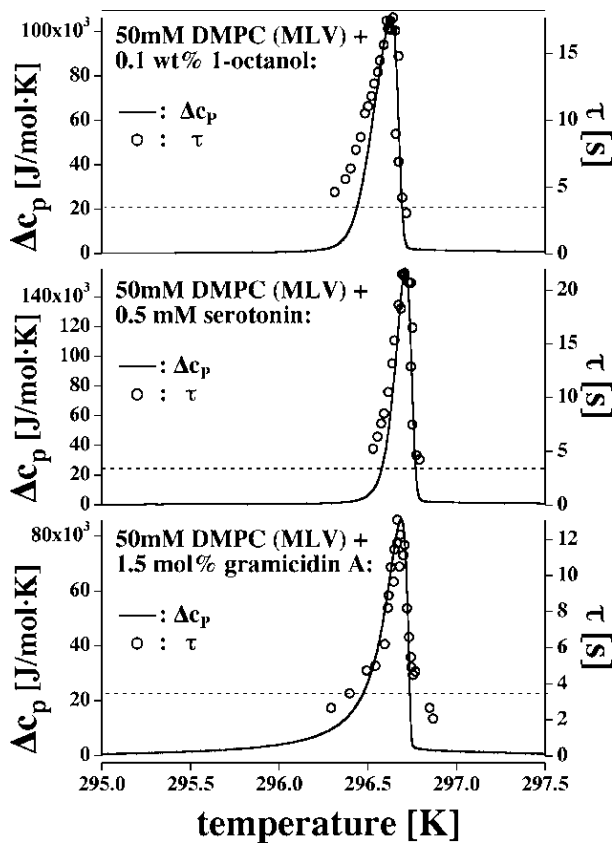}
	\parbox[c]{8cm}{ \caption{\textit{Comparisons between heat
	capacity profiles (\emph{solid curves}) and relaxation times
	(\emph{open circles}) of different multilamellar DMPC systems.
	Top: $50\: mM$ DMPC plus $0.1\: wt\: \%$ 1-octanol.  Center:
	$50\: mM$ DMPC plus $0.5\: mM$ serotonin.  Bottom: $50\: mM$
	DMPC plus $1.5\: mM$ gramicidin.  The drugs alter the
	temperature dependence of the heat capacity and the magnitude
	and temperature dependence of the relaxation times.  The
	dashed line indicates the resolution limit of the experiment
	(3-4 seconds).  }
	\label{fig:fig4}}}
    \end{center}
\end{figure}

In the following we investigated in which way other small molecules
and peptides influence the relaxation behavior of lipid membranes.
Therefore, we performed experiments ad\-ding different concentrations of
the anesthetic 1-octanol, the neurotransmitter serotonin and the
antibiotic gramicidin A to the DMPC lipid membranes.  As seen in
fig.~\ref{fig:fig2} these molecules broadened the melting transition
regime and shifted it to lower temperatures.  This has also an
influence on the relaxation behavior of the lipid membranes.  In
fig.~\ref{fig:fig4} measurements on the three different systems are
displayed.  Heat capacity profiles are taken from fig.~\ref{fig:fig2}.
Results were obtained from analyzing the calorimetric response after
jumps of $\pm 40 bar$.  Results from negative and positive pressure
jumps are not indicated by different symbols.  The dashed line always
indicates the time resolution of the setup.  Both heat capacity and
relaxation time scale of the different panels are different.  In all
cases we found a proportionality between heat capacity and relaxation
times.  Phenomenological constants were calculated to be
$(6.3\pm0.3)\cdot 10^8\: J\: K/mol\: s$ (DMPC plus $0.1\:$ wt \%
1-octanol), $(6.5\pm0.5)\cdot 10^8\: J\: K/mol\: s$ (DMPC plus $0.5\:$
mM serotonin) and $(5.8\pm0.5)\cdot 10^8\: J\: K/mol\: s$ (DMPC plus
$1.5\:$ mM gramicidin A), i.e. within error L is independent of the
system.  Note that deviations close to 3-4 seconds are most likely a
consequence of the finite time resolution of the experiment.

A series of PPC experiments were performed using different
concentrations of the respective molecules (\emph{data not shown}).
The maximum relaxation times obtained and the phenomenological
constants determined are displayed in
table~\ref{tab:relaxti_pheno_tab1}.  As it is seen in
fig.~\ref{fig:fig2} the addition of the chosen small drugs led to a
broadening of the heat capacity curves and to a decrease in the
maximum heat capacity value.  The maximum relaxation time was
influenced in the same way as is the maximum heat capacity value.  The
whole relaxation time profile behaves the same as does the heat
capacity curve.  The addition of molecules alter the relaxation
behavior of lipid domain formation in a systematic way.
\begin{figure}[htb!]
    \begin{center}
	\includegraphics[width=8cm]{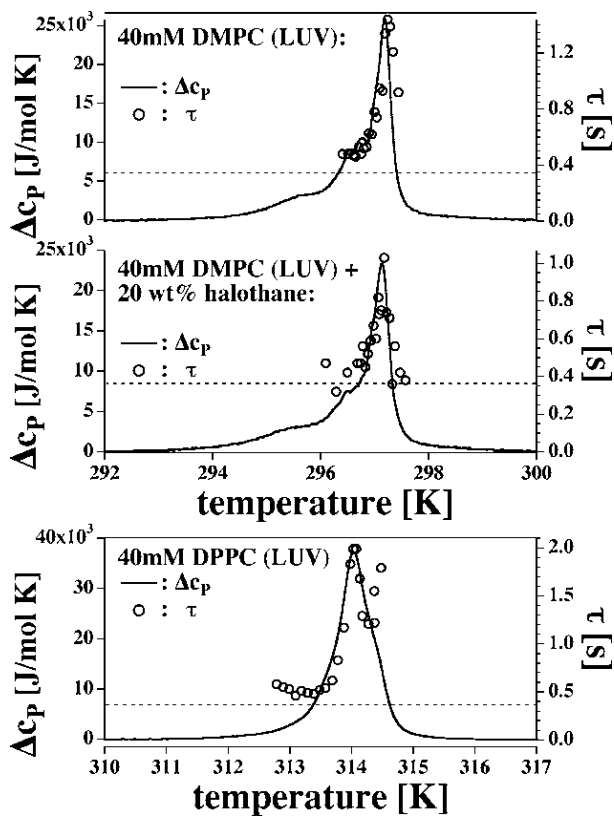}
	\parbox[c]{8cm}{ \caption{\textit{Heat capacity profiles
	(solid lines) and relaxation times (symbols) of three
	different large unilamellar vesicle systems.  Relaxation times
	were recorded with a thin thermocouple.  Top $40\: mM$ DMPC.
	Center: $40\: mM$ DMPC plus $20\: wt\: \%$ halothane.  Bottom:
	$40\: mM$ DPPC. Again a proportionality between heat capacity
	and relaxation times was found.  Halothane acts similar on the
	lipid membrane as 1-octanol.  The dashed line indicates the
	resolution limit of the experiment (around 350 milliseconds).
	}
	\label{fig:fig5}}}
    \end{center}
\end{figure}

Further experiments were conducted on solutions with lar\-ge unilamellar
vesicles.  We investigated LUVs of DMPC lipids with and without
addition of the anesthetics halothane.  We also studied LUVs of DPPC.
The respective heat capacity values and relaxation times are plotted
in fig.~\ref{fig:fig5}.  Relaxation times were obtained using a
thermocouple inserted into the pressure cell using positive pressure
jumps.  The dashed line represents the time resolution of the setup.
In all cases we found again that heat capacity and relaxation times
are proportional to each other.  Halothane has a similar effect on the
melting behavior as the anesthetic octanol.  It broadened and shifted
the heat capacity curve to lower temperatures.  The maximum relaxation
time was again decreased compared to that of the DMPC LUVs without
halothane.  The phenomenological constants for the LUV systems are
somewhat higher than those of the MLVs (see table~\ref{fig:fig5}).
The phenomenological constants obtained from measurements on MLVs
suggest that there is a constant independent of the lipid membrane
composition.  The same is true for LUVs, while we calculated different
phenomenological constants for MLVs and LUVs.  From our results we
calculated averaged phenomenological constants of $(6.7\pm 0.7)\cdot
10^8\: J\: K/mol\: s$ (MLVs) and $(18.0\pm2.2)\cdot 10^8\: J\: K/mol\:
s$ (LUVs).  The reason for this is unknown.  One can speculate on
whether relaxation processes are influenced by volume changes of the
vesicles in the case of LUV or by bilayer-bilayer interactions in the
case of MLV.

\begin{figure*}[htb]
    \begin{center}
	\includegraphics[width=15cm]{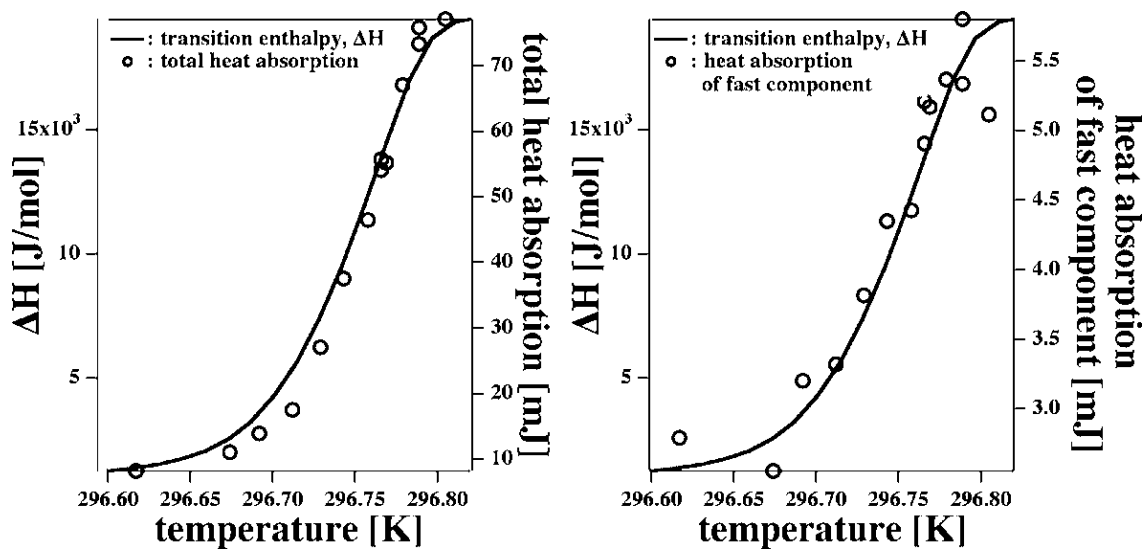}
	\parbox[c]{16cm}{ \caption{\textit{The solid curve in the left
	and right panel show the transition enthalpy of DMPC
	multilamellar vesicles as obtained from integrating the excess
	heat capacity profile.  The total heat compensation of the
	calorimeter after a pressure jump from below the transition to
	a given point within the transition is proportional to the
	transition enthalpy (\emph{circles, left panel}).  The signal
	typically contains one single-exponential slow component and a
	minor contribution from fast processes that show up as a
	signal with the characteristic time constant of the
	calorimetric setup.  The contribution from the fast processes
	(\emph{circles, right panel}) also displays a proportionality
	to the integrated heat.  This means that there is at least one
	relaxation processes faster than the time resolution of our
	experiment.  Its contribution, however, is less than 8\% of
	the overall signal.  Thus, the relaxation process of the
	cooperative events is well approximated by a single relaxation
	process. }
	\label{fig:fig6}}}
    \end{center}
\end{figure*}

In the literature up to five relaxation processes in the time range
from $ns$ to $s$ were reported for one lipid component systems
\cite{holzwarth:structdyn_PCmembr}.  In this study we assumed only one
relaxation time, which describes domain formation processes.  In order
to test whether there are contributions from faster relaxation
processes we analyzed the heat absorption after negative pressure
jumps as a function of temperature in comparison to the transition
enthalpy obtained from scanning calorimetry.  The transition enthalpy
of the DMPC (MLV) solution (see the heat capacity profile in
fig.~\ref{fig:fig2}) is shown as a solid curve in the left and right
panel of fig.~\ref{fig:fig6}.  The total heat absorption is displayed
in the left panel of fig.~\ref{fig:fig6}.  It is proportional to the
transition enthalpy.  All relaxation data contained contributions of a
fast component with a time scale corresponding to the instrument
response time.  The heat absorption due to this component also showed
a proportionality to the transition enthalpy (\emph{right panel of
fig.~\ref{fig:fig6}}).  This means that at least one faster relaxation
process is present which occurs on a time scale faster than our
experimental time resolution.  Analyzing all measurements we found
that the contribution of faster relaxation processes is lower than
$10\: \%$ of the total heat absorption.  Using the thermocouple signal
as a detection method we did not detect faster relaxation processes.
\begin{table}[h]
\begin{center}
    \footnotesize
\begin{tabular}{|l l l |}
\hline
Lipid Membrane System & $\tau_m/s$ & $L/ (10^8\frac{J\: K}{mol\: s})$
\\
\hline
 DMPC (MLV) & $31.0\pm1.5$ & \, $6.6\pm0.5$\\ 
+ $0.1\: wt\%$ 1-octanol (MLV) & $17.7\pm0.4$& \, $6.3\pm0.3$\\ 
+ $0.5\: wt\%$ 1-octanol (MLV) & $14.8\pm0.2$& \, $6.3\pm0.3$\\ 
+ $0.5\: mM$ serotonin (MLV) & $21.6\pm1.2$& \, $6.5\pm0.5$\\ 
+ $3.0\: mM$ serotonin (MLV) & \, $9.0\pm0.9$& \, $7.5\pm0.8$\\ 
+ $0.5\: mol\%$ gramicidin A (MLV)& $20.8\pm0.8$& \, $7.9\pm0.5$\\ 
+ $1.5\: mol\%$ gramicidin A (MLV)&  $13.1\pm0.8$& \, $5.8\pm0.5$\\
\hline
DMPC (MLV) & $25.2\pm0.1$ & \, $6.8\pm0.3$\\ 
DMPC (LUV) & \, $1.5\pm0.2$& $15.7\pm3.0$\\ 
+ $20\: wt\%$ halothane (LUV) & \, $1.0\pm0.1$& $20.0\pm1.0$ \\
DPPC (LUV) & \, $2.0\pm0.1$& $18.4\pm1.0$ \\ \hline
\end{tabular}
\parbox[c]{8cm}{\caption[Relaxation Times and Phenomenological
Constants]{\textit{Given are the relaxation times and phenomenological
constants at the transition midpoint of the studied systems.  In the
upper parts relaxation times were determined from the calorimeter
response.  In the lower part they were obtained from the temperature
change curve in the sample capillary.}}}
\label{tab:relaxti_pheno_tab1}
\end{center}
\end{table}


In total we can state that domain formation processes are related to
fluctuations in enthalpy at large excess heat capacities.  Relaxation
times and heat capacity are linear functions (within error).  This
linear relation is still true in the presence of drugs.  However, the
temperature dependence of both heat capacity and relaxation times are
altered by drugs in a simple systematic manner.  Fast relaxation
processes contribute less than 10\% to the overall heat.  Thus, they
probably do not represent the fluctuations in domain size linked to
the lipid chain state.

\section*{Discussion}
The melting of lipid membranes is a cooperative event involving many
lipid molecules.  Within the melting transition one finds domain
formation, but also changes in the elastic constants.  If pressure,
temperature or any other relevant intensive thermodynamic variable is
slightly changed, the lipid system changes its state and domains grow
or shrink while absorbing or releasing significant amounts of heat.
Thus, the domain formation process can be followed by monitoring the
heat change.  If we talk about relaxation times within the lipid
melting transitions we therefore basically refer to the time scale of
domain growth.  These processes have been studied by some groups
before using calorimetry \cite{vanOsdol1989, vanOsdol1991a,
vanOsdol1991b, vanOsdol1992, Grabitz2002}. One typically finds slow 
relaxation in transitions. This phenomenon is known as `critical 
slowing-down'.

Studies of relaxation processes in artificial membranes by use of
pressure perturbation calorimetry have been performed by our group
before \cite{Grabitz2002}.  We used the same and a refined version of
the setup as described in \cite{Grabitz2002} to extend the previously
published studies.  Relaxation processes in LUVs and the influence of
small drugs on the relaxation behavior of lipid domain formation in
LUVs and MLVs were investigated.  The typical time scales of domain
formation processes are related to cooperative fluctuations in
enthalpy and in all systems we found a proportionality between heat
capacity and relaxation times.  The relaxation processes depend on the
final state rather than on the initial state.  A remarkable finding is
that the drugs used modulate the time-scale of domain formation
processes in a similar manner as they influence the heat capacity
profiles.  They tend to broaden both heat capacity profiles and
relaxation time profiles.  After addition of anesthetics,
neurotransmitters and antibiotics the maximum relaxation time is
decreased in the same way as is the maximum heat capacity value .
Therefore, these drugs alter static and kinetic aspects of domain
formation.

Our findings support the theoretical prediction from Grabitz et al.
\cite{Grabitz2002} that heat capacity and relaxation time display a
linear relation not only for pure lipid membranes, but also for
membrane mixtures and mixtures with drugs that associate with 
membranes. The proportional constant between relaxation time and heat 
capacity takes the form $T^2/L$, where L is an phenomenological 
coefficient originating from Onsager's phenomenological equations.
We calculated average phenomenological constants to be $(6.7\pm
0.7)\cdot 10^8\: J\: K/mol\: s$ for MLVs and $(18.0\pm2.2)\cdot
10^8\: J\: K/mol\: s$ for LUVs made of DMPC and DPPC. The
constant L is different for MLV and LUV by about a factor of 2.5.
Surprisingly, however, the addition of drugs does not change the
phenomenological constant for a given vesicular system even though
they have significant influence on the heat capacity profiles.  With
measurements on suspensions of multilamellar DMPC vesicles we obtained
phenomenological constants of $6.6\cdot 10^8\: J\: K/mol\: s$
and $6.8\cdot 10^8\: J\: K/mol\: s$ using the two different
detection methods.  They agree within error.  Thus, the difference in
result does not seem to be related to the method of detection.  We do
not know the origin the difference between MLV and LUV. However, MLV
display much more cooperative transitions because of coupling between
adjacent membranes.  The fluctuations therefore contain a coupling in
the third dimension that might cause small differences in the
relaxation behavior.  The difference may also be related to the change
of vesicle size while changing the lipid membrane state and area.
This may involve permeation of water into the vesicles.

Anesthetics, neurotransmitters and antibiotics influence the melting
behavior of lipid membranes.  The anesthetic octanol shifts profiles
to lower temperatures.  Long chain alcohols (with chain length larger
than 10) shift them to higher temperatures
\cite{Tamura:highpressantagonism}.  The latter ones do not display
anesthetic potency.  Other anesthetics like halothane (this study and
unpublished data) or methoxyflurane \cite{trudell:anesthandpressure}
also lead to a decrease of the melting temperature.  It has been
stated that clinically relevant concentrations show effects on the
shift of melting temperatures \cite{trudell:anesthandpressure}.  The
effective potency of an anesthetic is correlated to its ability to
deplete melting transition temperatures
\cite{Kharakoz:phasetranssynexo}.  Serotonin possibly also acts as an
anesthetic \cite{Cantor:neurotransanesth} and influences the phase
melting transition.  This has also been shown for other
neurotransmitters, like GABA ($\gamma$-aminobutyric acid), dopamine,
caffeine and fluoxetin \cite{Polla:thermbiolmem}.  The interactions
with peptides or proteins and other small molecules like anesthetics
and neurotransmitters influence the domain structuring of lipid
membranes in the phase melting transition regime
\cite{seeger:fluct05}.  All these molecules influence relaxation
processes in a systematical way.

In previous studies kinetic aspects of melting transitions in
artificial single-component lipid membranes have been studied with a
whole variety of different techniques.  This study, however, is the
first one systematically investigating the effects of various small
drugs on relaxation processes in single lipid membranes and relating
it to the enhancement of fluctuations in the transition regime.  Van
Osdol and co-workers \cite{vanOsdol1992} studied the effect of the
anesthetics dibucaine on relaxation processes.  They claimed that even
though dibucaine clearly shows an influence on the melting behavior
only a small alteration of relaxation processes has been found.  This
finding does not agree with ours, where we have found that the drugs
influence relaxation processes  similar to their influence on
the heat capacity. 

In other studies relaxation times faster than the ones we measured and
up to five relaxation processes were reported
\cite{traeuble:schaltprozmemb, Tsong:kin_bilay, Tsong:relaxphen,
Elamrani:presjumrelax, Holzwarth:krittrueb, Genz:dynfluoresc,
holzwarth:structdyn_PCmembr, mitaku:ultrasonicrelax,
Halstenberg:CritFluct}.  Different spectroscopic detection methods
were used.  Previously, it has been discussed that some of the
experiments might have problems with a well-defined temperature, due
to the necessity of having windows for the detection.  These might act
as heat sinks \cite{Grabitz2002}.  In favor of longer relaxation times
are estimations using ac-calorimetry \cite{Yao:relax}.  The authors
found relaxation times faster than $120\: s$ (MLV of DMPC) and $260\:
s$ (MLV of DPPC).  They, however, did not give relaxation times in
dependence of temperature.  As mentioned, relaxation processes have
been reported from the $ns$ to minute regime.  One reason might lie in
an inaccurate temperature control in many optical methods.  However,
another reason, which seems likely to us is that the different
detection methods probe different relaxation processes.  In
\cite{holzwarth:structdyn_PCmembr} the different relaxation times
where assigned to relaxation processes reflecting chain properties
like the formation of kinks, free rotations of headgroups or the
formation of complex rotational isomers and to macroscopical
properties, i.e. the formation of clusters.  Our study does not
exclude the existence of several faster relaxation processes, but the
focus was only on one relaxation process, namely the formation of
domains which is responsible for most of the heat absorption.  We
found evidence that there is at least one faster relaxation process
which has a minor contribution to the total heat absorption (less than
10\%). Most previous studies agreed on a slowing down of relaxation processes
in the melting transition regime \cite{traeuble:schaltprozmemb,
Tsong:relaxphen, mitaku:ultrasonicrelax, Gruenewald:kininvest,
Elamrani:presjumrelax, holzwarth:structdyn_PCmembr, vanOsdol1989,
Grabitz2002, Halstenberg:CritFluct, vanOsdol1991a}.

Biological membranes also display melting transitions
\cite{melchior:transbiomembr, Jackson:EcoliDSC}.  From the knowledge
of the phenomenological constant and the excess heat capacity profile
were estimated relaxation times in the melting transition regime of
biological membranes \cite{Grabitz2002}.  The authors used a DSC
measurement of bovine lung surfactant from Ebel et al.
\cite{Ebel:enthvolchang} and a phenomenological constant similar to
the one for DMPC and DPPC MLV vesicles.  They estimated that
relaxation times in biological membranes would lie in a time regime of
up to $120\: ms$ in the transition regime.  We have calculated two
different phenomenological constants for MLVs and LUVs but the order
of magnitude is unaffected.  The time regime of these domain formation
processes is in a regime which seems to be important in biology.
These relaxation times do not only give the typical time scales of
domain formation processes, but also the life time of domains.
Domains are subject to fluctuations \cite{seeger:fluct05} and the
smaller they are the shorter their life time.  Very broad melting
profiles typically result in small domains.  In the ongoing discussion
on 'rafts' this may be of some relevance.  Their size is thought to be
smaller than 100 $nm$ \cite{Simons:liprafts} and therefore should they
display short life times.  Additional to their small size the short
life time may make it difficult to really detect them.

Domain formation processes in general have been argued to be important
in the control of biochemical reaction cascades in biological
membranes \cite{Melo:domainconnectionreactions, Vaz:phasetopol,
Thompson:domainstrucreaction, Hinderliter:controlsigntrans,
Salinas:changesenzymeactivity}.  Since drugs like anesthetics,
neurotransmitters and antibiotics also alter the lateral membrane
structure and relaxation times they might show an indirect action on
enzymatic processes.  For several enzymes it has been shown that the
existence of domains is necessary for their function
\cite{OpdenKamp:phospholipase74, OpdenKamp:phospholipase,
Gabriel:phospholipase, Lichtenberg:phospholipase,
Burack:LipBilayHeterophopholipase, Bolen:unstapkcactivation,
Dibble:laterheteroactpkc}.  In the case of phospholipase $A_2$ is was
stated that the magnitude of fluctuations determines the activity of
the enzyme \cite{Biltonen:statthermo}.  It may be easier for the
phospholipase to reach the target site at the lipid when the
fluctuations are large.  The kinetics of lateral membrane structure
might generally act as a control of biochemical reactions and their
time scales.

The action of general anesthetics is still not fully understood.  Many
groups presently favor a picture where anesthetics directly influence
the function of proteins \cite{Franks:genanaesth, Krasowski:genanesth,
Bovill:meyeroverton}.  Theories about the action of general
anesthetics mediated by the lipid membrane are put forward by several
other groups \cite{Cantor:genanesthetics, Cantor:meyeroverton,
Heimburg2007b, Heimburg2007c}.  The later models are in our opinion
more convincing because the action of anesthetics is known to be
linearly related to their solubility in lipid membranes, including the
noble gas Xenon that is inert and therefore unlikely to have specific
interactions with macromolecules.  Thinking about lipid-mediated
anesthesia, a possible implication of our findings might be related to
ion permeation through lipid bilayers.  The predicted relaxation time
of biological membranes in the several 10 ms range is just the typical
time scale of the opening and closing of ion channel proteins.
Recently, several groups started to argue that the lipid environment
has a direct influence on the channel opening statistics
\cite{Turnheim1999, Cannon2003, Schmidt2006}.  For this reason
anesthetics and neurotransmitters may display an indirect influence on
these proteins via their action on the lipid membrane.  This implies
that the channel lifetimes may be coupled to the relaxation time
scales of the lipid membranes, in particular since they are obviously
of right order.  It was also reported that permeation through pure
lipid bilayers in the absence of proteins is enhanced in the melting
transition regime and reaches a maximum at a temperature of maximum
heat capacity.  This finding was related to the magnitude of
fluctuations \cite{papa:permeab, Nagle:permeab,
Cruzeiro:permeability}.  Measurements on black lipid membranes in the
melting transition regime yielded characteristic fluctuations of the
transmembrane ion currents
\cite{Antonov:singleionchannelslipidbilayer,
Antonov:perforationlipidmembr, Kaufmann:ionchannels}.  No such
currents were found for lipid membranes at temperatures well above or
below the transition midpoint temperature of one component lipid
membranes \cite{Antonov:perforationlipidmembr}.  The life times of
these lipid pores within the phase transition regime were found to be
as long as several seconds, in agreement with the relaxation time
scales found by us.\\

In this study we have shown that relaxation times of domain formation
processes in simple model systems are proportional to the excess heat
capacity.  Relaxation processes slow down in the melting regime.
Several drugs studied by us broaden the melting transition profiles
and shift them to lower temperatures.  This influences the relaxation
behavior in a simple manner.  This is presumably true for all drugs,
peptides and proteins influencing the melting behavior of lipid
membranes.  Clearly, the free energy of membranes includes the
chemical potentials of all membrane associated molecules and it is
therefore just a consequence of thermodynamics that the functions of
states change in a coherent manner due to the change in one of these
variables.  This is likely to be of great importance for the general
understanding of the function of biological membranes.  \\


\textbf{Acknowledgments}

HM.S. was supported by the Deutsche Forschungsgemeinschaft (DFG;
grant HE1829/11-1).


\end{document}